\documentclass[10pt,a4paper]{article}
\usepackage[utf8]{inputenc}
\usepackage{amsmath}
\usepackage{amsfonts}
\usepackage{amssymb}
\usepackage{graphicx}
\usepackage{algorithm}
\usepackage{algorithmic}
\usepackage[left=2.50cm, right=2.50cm, top=2.50cm, bottom=2.50cm]{geometry}
\begin{document}

\title{Steganography and Broadcasting}
\author{Fabrice P. Tachago, Stephane G. R. Ekodeck, Ren\'{e} Ndoundam \\
{\small University of Yaounde I, LIRIMA, Team GRIMCAPE, P.o.Box 812 Yaounde, Cameroon} \\
{\small IRD, UMI 209, UMMISCO, IRD France Nord, F-93143, Bondy, France; } \\
{\small Sorbonne Unversit\'es, Univ. Paris 06, UMI 209, UMMISCO, F-75005, Paris, France } \\
{\small E.mail : tachagofabrice@gmail.com, ekodeckstephane@gmail.com, ndoundam@yahoo.com } }
\date{}
\maketitle{}

\begin{abstract}
Informally, steganography is the process of exchanging a secret message between two communicating entities so that an eavesdropper may not know that a message has been sent. After a review of some steganographic systems, we found that these systems have some defects. First, there are situations in which some concealment algorithms do not properly hide a secret message. Second, to conceal one bit of a secret message, some ask at least five documents and make at least two sampling operations, thus increasing their run-times. Considering the different ways to communicate with the receiver, we propose two steganographic systems adapted to the email communication whose algorithms are deterministic. To hide one bit of a secret message, our steganographic systems need only one document and performs one sampling operation and therefore significantly reduces the run-time.

\end{abstract}
\textit{Keywords}: steganography, run-time, email

\section{Introduction}
Steganography is the art of concealing secret messages in seemingly innocent media. The use of steganography is not new, as it dates back to antiquity \cite{key-8,key-9,key-10}. But scientific study began when G. J. Simmons \cite{key-1} formulated the problem of communication between two prisoners, which can be stated as follows: Alice and Bob, both prisoners, held in separated cells and remote from one another wish to establish an escape plan without Eve the guardian knowing about it. With Eve allowing them to communicate on condition that the exchanged messages are clear and understandable.

\paragraph{State of art:} Two formalisms have been proposed to solve this problem: Theory of Information and Theory of Complexity. In the theory of information, C. Cachin \cite{key-2} provided a steganographic protocol based on a probabilistic model. It defines the security of a steganographic protocol by a relative entropy between the distribution of covertexts and stegotexts. Here, a protocol is said to be perfectly safe if the relative entropy is null.
In the theory of complexity approach, N. Hopper, J. Langford and L. Von Ahn \cite{key-3}, offered a rigorous formalization of a protocol whose security is based on the difficulty for an adversary to distinguish the distribution of covertexts and stegotexts in polynomial time.

N. Hopper and al. \cite{key-3} proposed a protocol based on a function called rejection sampling. The aim of function is to find in a distribution of covertexts, a covertext which its image by a given pseudo-random function is equal to the secret message's bit that we want to hide. We present now the algorithms proposed in \cite{key-3}.

\begin{algorithm}[h]
\textbf{Procedure} $RS$ 
	\caption{ Rejection sampling\cite{key-3}}
	\label{sampling rejection of the system 1}
	\begin{algorithmic}[1]
		\STATE \textbf{Input}: \textit{Key} $K1$, \textit{Target} $x$, \textit{iteration} $count$, \textit{history} $h$\\
		\STATE $i = 0$\\
		\REPEAT
		\STATE $ c \leftarrow \mathbf{S}(h)$ \\
		\STATE \textbf{Increment} $i$ \\
		\UNTIL $i = count$ or $F_{K1}(c) = x$ \\
		\STATE \textbf{Output:} $c$
	\end{algorithmic}

\end{algorithm}	
	
\begin{algorithm}[h]
\textbf{Procedure} $\mathcal{S}1.Embed$ 

	\begin{algorithmic}[1]
		\STATE \textbf{Input}: \textit{Key} $K$, \textit{hiddentext} $m \in \lbrace 0,1 \rbrace ^{*}$, \textit{history} $h$ 
		\STATE \textbf{Parse} $m$ \textbf{as} $m_{1}^{1}||m_{2}^{1}||\cdots||m_{n}^{1} $
		\FOR{$ i = 1 \cdots n$} 
		\STATE $c_{i} = RS(K,m_{i},\vert K \vert,h)$ 
		\STATE $h = h||c_{i}$ 
		\ENDFOR
		\STATE \textbf{Output:} $c_{1}c_{2}\cdots c_{n} $ 	
	\end{algorithmic}

\caption{Embedding procedure \cite{key-3}}
\label{Embedding procedure of the system 1}
\end{algorithm}

\begin{algorithm}[h]
\textbf{Procedure} $\mathcal{S}1.Extract$ 
	\begin{algorithmic}[1]
		\STATE \textbf{Input:} \textit{Key} $K$, \textit{stegotexts} $c_{1}c_{2}\cdots c_{n}$ 
		\FOR{$ i = 1 \cdots n$ } 
		\STATE $m_{i} = F_{K}(c_{i})$ 
		\ENDFOR
		\STATE $m = m_{1}||m_{2}||\cdots||m_{n}$
		\STATE \textbf{Output:} $m$ 
	\end{algorithmic}

\caption{Extracting procedure \cite{key-3}}
\label{Extracting procedure of the system 1}
\end{algorithm}

The concealment algorithm applies the $RS$ procedure for each bit. The extracting algorithm simply applies pseudo-random function to each document received.\\
This system possesses two defects: firstly, it needs more sampling operations (expensive operation \cite{key-7}) which increases its run-time. The algorithm runs in $\mathit{O}(\vert K \vert n) $. Secondly this system is not safe for a covertexts distribution which owns a small min-entropy \cite{key-6}.

To correct these flaws, Hopper and al. \cite{key-4} have proposed another protocol. That protocol is still based on a rejection sampling function. But this time, it makes at most two sampling operations and uses a error correcting code to increase the reliability of the steganographic system.

\begin{algorithm}[h]
	\textbf{Procedure} $RS$
	\begin{algorithmic}[1]		
		\STATE \textbf{Input}: \textit{Key} $K1$, \textit{Synchronized} $N1$, \textit{Target} $x$, \textit{iteration} $count$, \textit{history} $h$
		\STATE $i = 0$\\
		\REPEAT
		\STATE $c \leftarrow \mathbf{S}(h)$ 
		\STATE\textbf{Increment} $i$ 
		\UNTIL{$i = count$ or $F_{K1}(N1,c) = x$ } 
		\STATE \textbf{Output}: $c$
	\end{algorithmic}
	\caption{rejection sampling \cite{key-4}}
	\label{rejection sampling of the system 2}
\end{algorithm}		

\begin{algorithm}[H]
\textbf{Procedure} $\mathcal{S}2.Embed$ 

\begin{algorithmic}[1]
	\STATE \textbf{Input}: \textit{Key} $K$, \textit{Synchronized} $N$, \textit{hiddentext} $m' \in \lbrace 0,1 \rbrace ^{*}$, \textit{history} $h$ 
	\STATE \textbf{Let} $ m = Enc(m') $ 
	\STATE\textbf{Parse} $m$ \textbf{as} $m_{1}^{1}||m_{2}^{1}||\cdots||m_{n}^{1} $
	\FOR{$ i = 1 \cdots n$}
	\STATE $c_{i} = RS(K,N,m_{i},2,h)$ 
	\STATE $h = h||c_{i}$ 
	\STATE \textbf{Increment} $N$ 
	\ENDFOR
	\STATE \textbf{Output:} $c_{1}c_{2}\cdots c_{n} $
\end{algorithmic}

\caption{Embedding procedure \cite{key-4}}
\label{Embedding procedure of system 2}
\end{algorithm}

\begin{algorithm}[H]
\textbf{Procedure} $\mathcal{S}2.Extract$ 
	\begin{algorithmic}[1]
		\STATE \textbf{Input}: \textit{Key} $K$, \textit{Synchronized} $N$, \textit{stegotexts} $c_{1}c_{2}\cdots c_{n}$ 
		\FOR{$ i = 1 \cdots n$} 
		\STATE $\tilde{m_{i}} = F_{K}(N,c_{i})$
		\STATE \textbf{Increment} $N$
		\ENDFOR
		\STATE $\tilde{m} = \tilde{m_{1}}||\tilde{m_{2}}||\cdots||\tilde{m_{n}}$\\
		\STATE \textbf{Output:} $Dec(\tilde{m})$ 	
	\end{algorithmic}

\caption{Extracting procedure \cite{key-4}}
\label{Extracting procedure of system 2}
\end{algorithm}

The use of a error correcting code is not only time expensive but reduces the transmission rate of the steganographic system (number of bits of secret message transmitted by covertext sent). Indeed, without use of error correcting code the failure probability of concealment of this system varies between 1/4 and 1/3. It depends on the covertexts distribution.

Shannon has shown that for a channel having a probability $p$ of distortion of a symbol, the capacity of this channel is equal to $ 1 - H(p)$. On such a channel, one can reliably communicate with a corrector code with a rate close to $1 - H (p)$. Taking p = 1/4, the rate goes to 0.2 so covertexts 5 for a single secret message bit transmitted. their concealment algorithm runs in $\mathit{O}(\frac{1}{1 - H (p)} \times n^{'}) $ plus time of error correcting code where $n^{'}$ is the length of secret message.

L. Reyzin and S. Russell \cite{key-6} generalized the protocol proposed in \cite{key-3} in order to be safe for steganographic distribution which has a small min-entropy. L. Reyzin and al. proceeded as follows: to hide a bit of secret message, it uses $t$ covertexts instead of only one. By doing so, it increases the min-entropy of the covertexts distribution. his concealment algorithm runs in $\mathit{O}(\vert K \vert \times t \times n) $.

\begin{algorithm}
	\textbf{Procedure} $RS$
	
	\begin{algorithmic}[1]
		\STATE \textbf{Input}: \textit{Key} $K1$, \textit{Target} $y$, \textit{iteration} $count$, \textit{history} $h$, \textit{number of covertexts} $t$
		\STATE $i = 0$
		\REPEAT
		\STATE $h' = h$
		\FOR{$j = 1 \cdots t$} 
		\STATE $x_{j} \leftarrow \mathbf{S}(h)$ 
		\STATE$h' = h'||x_{j}$\\
		\ENDFOR
		\STATE $x = x_{1}x_{2}\cdots x_{t} $
		\STATE \textbf{Increment} $i$ 
		\UNTIL{$i = count$ or $F_{K1}(x) = y$} 
		\STATE \textbf{Output:} $x$
	\end{algorithmic}
	
	\caption{Rejection sampling \cite{key-6}}
	\label{Rejection sampling of system 3}
\end{algorithm}	
	
\begin{algorithm}[h]
\textbf{Procedure} $\mathcal{S}3.Embed$ 

	\begin{algorithmic}[1]
		\STATE \textbf{Input}: \textit{Key} $K$, \textit{hiddentext} $m \in \lbrace 0,1 \rbrace ^{*}$, \textit{history} $h$, \textit{number of covertexts} $t$ 
		\STATE \textbf{Parse} $m$ \textbf{as} $m_{1}^{1}||m_{2}^{1}||\cdots||m_{n}^{1} $
		\FOR{$ i = 1 \cdots n$} 
		\STATE $c_{i} = RS(K,m_{i},\vert K \vert,h,t)$ 
		\STATE $h = h||c_{i}$ \\
		\ENDFOR
		\STATE \textbf{Output:} $c_{1}c_{2}\cdots c_{n} $ 
	\end{algorithmic}

\caption{Embedding procedure \cite{key-6}}
\label{Embedding procedure of the system 3}
\end{algorithm}

\begin{algorithm}[H]
\textbf{Procedure} $\mathcal{S}3.Extract$ 
	\begin{algorithmic}[1]
		\STATE \textbf{Input}: \textit{Key} $K$, \textit{stegotexts} $c_{1}c_{2}\cdots c_{n}$ 
		\FOR{$ i = 1 \cdots n$} 
		\STATE $m_{i} = F_{K}(c_{i})$ 
		\ENDFOR
		\STATE $m = m_{1}||m_{2}||\cdots||m_{n}$\\
		\STATE \textbf{Output:} $m$ 
	\end{algorithmic}

\caption{Extracting procedure \cite{key-6}}
\label{Extracting procedure of the system 3}
\end{algorithm}

N. Hopper et al. \cite{key-5} proposed another steganographic system that no longer uses error correcting code. To hide a single bit, this system makes $t$ copies of the secret message. And for each copy, it seeks a covertext in the channel whose image via a pseudo-random function is equal to the secret message by taking more than twice from the channel. the concealment and extracting algorithms run in $\mathit{O}(t) $ for one bit. 

\begin{algorithm}[h]
\textbf{Procedure} $\mathcal{S}4.Embed$ 
	\begin{algorithmic}[1]
		\STATE \textbf{Input}: \textit{Key} $K$, \textit{hiddentext} $m \in \lbrace 0,1 \rbrace $, \textit{history} $h$, \textit{Synchronized} $N$ 
		\FOR{$ i = 1 \cdots t$} 
		\STATE $d_{i},d^{'}_{i} \leftarrow \mathbf{S}(h)$
		\IF{$F_{K}(N + i,d_{i}) = m$} 
		\STATE $s_{i} = d_{i}$
		\ELSE
		\STATE $s_{i} = d_{i}^{'}$
		\ENDIF
		\STATE $h = h||s_{i}$ 
		\ENDFOR
		\STATE $N = N + t$\\
		\STATE \textbf{Output:} $s_{1}\cdots s_{t} $ 
	\end{algorithmic}

\caption{Embedding procedure \cite{key-5}}
\label{Procedure de dissimulation du systeme 4 }
\end{algorithm}

\begin{algorithm}[h]
\textbf{Procedure} $\mathcal{S}4.Extract$ 
	\begin{algorithmic}[1]
		\STATE \textbf{Input:} \textit{Key} $K$, \textit{Synchronized} $N$, \textit{stegotexts} $s_{1}\cdots s_{t}$ 
		\STATE $C = 0$
		\FOR{$ i = 1 \cdots t$ } 
		\STATE $C = C + F_{K}(N,s_{i})$ 
		\STATE \textbf{Increment} $N$
		\ENDFOR
		\IF{$C > t/2$} 
		\STATE $m = 1 $
		\ELSE
		\STATE $m = 0 $ 	
		\ENDIF	
		\STATE \textbf{Output}: $m$ 
	\end{algorithmic}

\caption{Extracting procedure \cite{key-5}}
\label{La procedure extraction du systeme 4}
\end{algorithm}

\section{Our contribution}
Hopper et al. \cite{key-3} proposed a steganographic system whose transmission rate is 1 bit per document sent. In order to improve the safety of the system to expand the channel distributions that have small min-entropy, they proposed in \cite{key-4} a system that reduces the transmission rate to 1 bit for 5 sent documents. Generalizing the system \cite{key-3} to be applied to channels that have small min-entropy, Reyzin L. and S. Russell \cite{key-6} propose a system which also reduces the transmission rate to $1$ bit for $t$ sent documents. Hopper et al. \cite{key-5} proposed a system that has the same transmission rate as \cite{key-6}.

The systems proposed in \cite{key-4,key-5,key-6} significantly reduce the transmission rate of one of [3]. These systems preserve probabilistic property of the concealment algorithm \cite{key-3} and ask several sampling operations. The run-time of all these concealment algorithms is considerable: generally $\mathit{O}(\alpha .n) $.

We propose two steganographic systems for the communication by email with concealment algorithms are deterministic. To hide a secret message bit, our algorithms only perform one sampling operation and transmission rate is 1 bit document sent by reducing their execution time in $\mathit{O}(n) $. The safety of protocols is reduced to that of a pseudo-random function.

This article is organized as follows: In section 3, we give some definitions that support the solution developed in this work. In section 4, we present our steganographic protocols and we conclude in section 5.

\section{Definitions}
\paragraph{Formatting convention:} Let $<a_{1},a_{2},\cdots,a_{n}>$ an array of elements. Two tables are said to be equal if the same index items are equals.

\subsection{Channel}
Let $\sum$ be a set of documents. A channel $\mathcal{C}$ is a function that takes as input a history $h \in \sum^{*}$ and provides a distribution probability $\mathcal{D}_{h} $.\\
The channel formalizes a normal communication between two entities. We define here a normal communication by email making use of the channel.

Let $\mathbb{A} $ be a set of array of addresses and $\mathcal{A}$ a uniform distribution on $\mathbb{A} $. We define an email by a triplet $ (d, adr, t)$ where $d \in \sum$, $t$ the sent date of message and $adr \in \mathbb{A}$ an array of addresses containing the receivers' addresses.\\
Let $h= d_{1}d_{2}\cdots d_{n} $ be the history of already sent messages. A mail $ (d, adr, t)$ is said to be normal if $d$ is drawn at random from the channel and $adr$ is chosen at random in $\mathcal{A} $.

\subsection{Steganographic system}
Let $\mathbb{A} $ be a set of array of addresses and $\sum$ a set of documents. A steganographic protocol or steganographic system is a pair of algorithms $\mathcal{S} = (\mathcal{S}.Embed,\mathcal{S}.Extract)$: 
	\begin{itemize}
		\item \textit{$\mathcal{S}.Embed$ takes as input a key $K \in \lbrace 0,1 \rbrace ^{k}$, a string $m \in \lbrace 0,1 \rbrace ^{*}$ (hiddentext), a history $h$ and has access to $\mathbf{S}$. $\mathcal{S}.Embed(k,m,h)$ returns a sequence of mails $s_{1}s_{2}\cdots s_{n}$ where $s_{i} = (d_{i},adr_{i},t_{i})$.}
		\item \textit{$\mathcal{S}.Extract$ takes as input a key $K$, two sequences of messages $s_{1}s_{2}\cdots s_{n_{1}}$ and $s_{1}^{'}s_{2}^{'}\cdots s_{n_{2}}^{'}$. $\mathcal{S}.Extract$ returns the secret message $m$ }
	\end{itemize}
	
\subsubsection*{Reliability}
$\mathcal{S}.Embed$ and $\mathcal{S}.Extract$ must satisfy the following relationship:
\[\forall m : \Pr[\mathcal{S}.Extract(K,\mathcal{S}.Embed(K,m,h)) = m] = 1 \]

\subsection{Security}
Intuitively we require, according to the security of steganographic system, that no effective adversary $G$ can distinguish mails returned by the concealment algorithm from normal mails. We assume that $G$ knows distributions $\mathcal{D}_{h}$ and $\mathcal{A}$. We allow the adversary $G$ to have access to $\mathcal{S}.Embed$ and $h$ (communication history between Alice and Bob), and select a message $m$. Only the key is not known by $G$. We model an attack against a steganographic system as a game played by a passive adversary as: $G$ has access to the oracle $\mathbf{M}$ that can be either:
\begin{itemize}
	\item $ST$. The oracle $ST$ has a uniformly chosen key $K$ and responds to requests $(m, h)$ with a mail drawn sequence $\mathcal{S}.Embed(K, m, h)$.
	\item $CT$. The $CT$ oracle also has a uniformly chosen key $K$ and respond to requests $(m, h)$ with a sequence of normal mails $s_{1}s_{2}\cdots s_{n}$ where $s_{i} = (d_{i},adr_{i},t_{i})$, $ d_{i} \in \mathcal{D}_{hd_{1}d_{2}\ldots d_{i-1}}$ for $1\leq i \leq n$. $adr_{i}$ is randomly selected in $\mathcal{A}$. $n$ is the number of mails returned by $\mathcal{S}.Embed(K, m, h)$.
\end{itemize}
 
After interaction with its oracle, G sets out a bit which represents his assumption about the type of $\mathbf{M}$. He puts 1 to say that $\mathbf{M}$ is of type $ST$ and 0 otherwise. We define the advantage of $G$ against a steganographic system $\mathcal{S}$ for a channel $\mathcal{C}$ by:
\[ \textbf{Adv}_{\mathcal{S},\mathcal{C}}^{SS}(G)= \vert \Pr_{K \leftarrow \mathcal{K}}[G^{\mathbf{M} = ST} = 1] - \Pr_{K \leftarrow \mathcal{K}}[G^{\mathbf{M} = CT} = 1] \vert\]
where the probability is taken as the random effect of $ST$, $CT$ and the choice of $G$. We define the insecurity of $\mathcal{S}$ by:
\[ \textbf{InSec}_{\mathcal{S},\mathcal{C}}^{SS}(t,q,l) = \max_{G \in \mathcal{G}(t,q,l)} \lbrace \textbf{Adv}_{\mathcal{S},\mathcal{C}}^{SS}(G) \rbrace \]
where $\mathcal{G}(t,q,l)$ is the set of all the adversaries that makes at most $q$ queries to the oracle, totaling at most $l$ bits (hiddentexts) and runs at most in $t$ steps.

\subsection{Pseudorandom function}
Let $\mathcal{F} = \lbrace F_{K} \rbrace_{K \in \lbrace 0,1 \rbrace^{n}}$ be a family of functions all with the same domain and co-domain.
Let $A$ be a probabilistic adversary with access to $\mathbf{Fn} $ sampling oracle. The prf-advantage of $A$ to $\mathcal{F}$ is:
\[\textbf{Adv}_{F}^{prf}(A) = \vert
		\Pr_{K \leftarrow \lbrace 0,1 \rbrace^{n}}[A^{\mathbf{Fn} = F_{K}} = 1] - \Pr [A^{\mathbf{Fn} = f} = 1]\vert
	 \] 

where $f$ is a random function of the same domain and co-domain $F_{K}$. The insecurity of $\mathcal{F}$ is given by the following formula:
$$ \mathbf{InSec}_{F}^{prf}(t,q) = \max_{A \in \mathcal{A}(t,q)}\lbrace \textbf{Adv}_{F}^{prf}(A) \rbrace $$

where $\mathcal{A}(t,q)$ denotes the set of opponents performing at most $t$ steps and makes at most $q$ queries to the oracle.

\section{Our protocols}
We present in this section our steganographic protocols that conceal several bits. The security of these protocols is based on the difficulty to solve a cryptographic problem: the break of a pseudo-random function.

Each of these protocols uses two primitives: \textit{extractDocument} which takes as input a mail and returns the document in this email and \textit{extractAddresses} which takes as input a mail and returns an array of the destination email addresses.

We assume also that Bob owns two addresses: \textit{address1} and \textit{address2} known by Alice and Eve.

\subsection{Steganographic protocol 1}
\subsubsection{Secret steganographic state for one bit}

Alice and Bob share a channel $\mathcal{C}$. $\mathcal{F}$ is a pseudo-random function where $F_{K}: \lbrace 0,1 \rbrace ^{d} \times \sum \rightarrow \lbrace 0,1 \rbrace$. Alice and Bob possess a secret key $K \in \lbrace 0,1 \rbrace ^{k}$ and are synchronized by a counter $N \in \lbrace 0,1 \rbrace^{n}$. Let $\mathbb{A} = \lbrace <address1>, <address2> \rbrace$ be a set of array of addresses and $\mathcal{A} $ an uniform distribution of $\mathbb{A} $. The following algorithms allow Alice and Bob to hide and extract a bit of secret message.

\begin{algorithm}[H]
\textbf{Procedure} $\mathcal{S}.EmbedOneBit$ 
	\begin{algorithmic}[1]
		\STATE \textbf{Input}: \textit{Key} $K$, \textit{Hiddentext} $m \in \lbrace 0,1 \rbrace$, \textit{Synchronized} $N$, \textit{history} $h$, \textit{Sent date} $t$ 
		\STATE $ d \leftarrow \mathbf{S}(h)$ \\
		\IF{$(F_{K}(N,d) = m)$} 
		\STATE $s = (d,<address1>,t)$ 
		\ELSE
		\STATE $s = (d,<address2>,t)$ 
		\ENDIF
		\STATE \textbf{Increment} $N$ 
		\STATE $h = h||d$ \\
		\STATE \textbf{Output:} $s$
	\end{algorithmic}

\caption{Embedding procedure for one bit}
\label{Embedding procedure of the protocol for one bit}
\end{algorithm}

\begin{algorithm}[H]
\textbf{Procedure} $\mathcal{S}.ExtractOneBit$ 
	\begin{algorithmic}[1]
		\STATE \textbf{Input:} \textit{Key} $K$, \textit{Synchronized} $N$, \textit{mail} $s$ 
		\STATE $d =$ extractDocument($s$)
		\STATE $ m = F_{K}(N,d)$ 
		\IF{$(extractAddresses(s) = \hspace*{0.2cm} <address2>) $} 
		\STATE $m = \overline{m}$\\
		\ENDIF
		\STATE \textbf{Increment} $N$ \\
		\STATE \textbf{Output:} $m$ 
		
	\end{algorithmic}
\caption{Extracting procedure for one bit}
\label{Extracting procedure of the protocol for one bit}
\end{algorithm}

The idea behind this algorithm is simple. Suppose Alice and Bob can communicate using two envelopes of different colors (white and brown). Alice, to send a message to Bob, draws a document $d$ from the channel, evaluates it by the function $F_{K}(N,\cdot)$. If the result $F_{K}(N,d)$ is equal to the bit of message she wants to hide, Alice sends the document $d$ in the white envelope otherwise it's sent in the brown envelope. When Bob receives the document $d$ in a brown envelope, he takes the complement of the secret message provided by the document otherwise he takes the bit as provided.

\paragraph*{Run-time:}

%The run-time of this protocol is practically constant for concealment and extraction algorithms and depends on the size of the secret message that is obviously 1: $T = \mathit{O}(1)$.
The complexity in time is constant: $T = \mathit{O}(1)$.

\paragraph*{Reliability:}

Our protocol uses deterministic algorithms, therefore we can easily find all the bits embedded in the mails.

\paragraph*{Insecurity:}
We show here that the task of an adversary to distinguish, between the distribution of mails returned by the concealment algorithm from normal mails, is more difficult than an adversary of a pseudo-random function. The proof of the theorem below is similar to the ones of \cite{key-4,key-5}. \\
\\
First, we will show that the distribution of documents returned by the concealment algorithm is equal to that of the channel. Then we also show that when replacing the function $F_{K}(\cdot,\cdot) $ by a random function, the distribution of addresses returned by concealment algorithm is equal to $\mathcal{A} $. And later, we will build an opponent against a pseudorandom function that has the same advantage an adversary of a steganographic system.\\
\\
\textbf{Lemma 1:} \textit{The probability of a document received from channel, returned by $\mathcal{S}.EmbedOneBit(K, m, N, h)$ is equal to the probability of this document in the channel $\mathcal{D}_{h}$.}
\\
\\
\textbf{Proof:} Let $d_{i}$ document content in the email back $\mathcal{S}.EmbedOneBit(K, m, h)$. The probability that $d_{i}$ is put in output only depends on its drawn in the channel conditioned by history $h$. Thus

$$\Pr_{(d_{i},adr_{i},t_{i}) \leftarrow \mathcal{S}.EmbedOneBit(K, m,N, h)}[d_{i}] = \Pr_{\mathcal{D}_{h}}[d_{i}] $$
\\
\textbf{Lemma 2:} When the function $F_{K}(\cdot,\cdot) $ is replaced by a random function $f$, for all $h \in \sum^{*}$ the probability of address in the mail returned by $\mathcal{S}.EmbedOneBit(K, m, N, h)$ is equal to her probability in $\mathcal{A}$.
\\
\\
\textbf{Proof}: Let $m$ be the bit of secret message. The probability that sends a document to the address $<adresseI> \in \mathbb{A}$ depends on the evaluation of the document and the counter N in the function $f$. And as the counter is incremented every sent document the entrance $(N, d)$ will always be different for the opponent even if a document is sent more than once.
$$ \Pr[<adresseI>] = \Pr[f(N,d) = m]$$
 $$ \qquad = \frac{1}{2}$$
\\
\\
\textbf{Theorem 1}: $$ \textbf{InSec}_{\mathcal{S},\mathcal{C}}^{SS}(t,q,q) \leq \textbf{InSec}_{F}^{prf}(t + q\mathit{O}(1),q)$$
 \\
\textbf{Proof}: Let $G \in \mathcal{G}(t,q,l)$ an adversary for $\mathcal{S}$. We construct an opponent $A$ for a pseudorandom function $\mathcal{F} $ with same advantage $G$. The construction algorithm is given below:
\begin{itemize}
	\item $A^{\mathbf{Fn}}$ works by executing $G$ and intercepting all queries sends his oracle $\mathbf{M}$,
	\item To respond these demands, $A^{\mathbf{Fn}}$ simulates concealment algorithm: $\mathcal{S}.EmbedOneBit$ using the oracle $\mathbf{Fn}$ instead of $F_{K}$ and gets a result $c$; and starts with $G$ with c as response to its request.
	\item When $G$ stops with output $b$, $A$ also puts $b$ output.
\end{itemize}
 
 Clearly, where $\mathbf{Fn} = F_{K}$, $A^{\mathbf{Fn}}$ perfectly simulates the oracle $ST$ of $G$, as follows:
 \begin{equation}
 \Pr_{K \leftarrow \lbrace 0,1 \rbrace^{k}}[A^{\mathbf{Fn} = F_{K}} = 1] = \Pr_{K \leftarrow \mathcal{K}}[G^{\mathbf{M} = ST} = 1] 
 \end{equation}
 By the previous Lemma 1 and Lemma 2, when $\mathbf{Fn}$ is a random function $f$, $A^{\mathbf{Fn}}$ perfectly simulates the $CT$ oracle $G$:
 \begin{equation}
 \Pr[A^{\mathbf{Fn} = f} = 1] = \Pr_{K \leftarrow \mathcal{K}}[G^{\mathbf{M} = CT} = 1] 
 \end{equation}
 Subtracting (2) from (1) we obtain:
 $$ \textbf{Adv}_{F}^{prf}(A) = \textbf{Adv}_{\mathcal{S},\mathcal{C}}^{SS}(G)$$
 
 The theorem follows by definition of insecurity. 
 %$$ \textbf{InSec}_{\mathcal{S},\mathcal{C}}^{SS}(t,q,q) \leq \textbf{InSec}_{F}^{prf}(t + q \mathit{O}(1),q)$$

 \subsubsection{Secret steganography state for several bits}
 We now present our first protocol that allows steganography to conceal several bits. The algorithms below allow Alice and Bob to embed and extract more bits of secret messages.
 
 \begin{algorithm}[h]
 \textbf{Procedure} $\mathcal{S}.Embed$ 
 	\begin{algorithmic}[1]
		\STATE \textbf{Input}: \textit{Key} $K$, \textit{Hiddentext} $m \in \lbrace 0,1 \rbrace ^{*}$,\textit{Synchronized} $N$, \textit{history} $h$, \textit{Sent date} $t$ 
		 \STATE \textbf{Parse} $m$ \textbf{as} $m_{1}^{1}||m_{2}^{1}||\cdots||m_{n}^{1} $
		 \FOR{$ i = 1 \cdots n$} 
		 \STATE $s_{i} = \mathcal{S}.EmbedOneBit(K,m_{i},N,h,t)$ 
		 \STATE $d_{i} = extractDocument(s_{i})$
		 \STATE $h = h||d_{i}$ 
		 \STATE \textbf{Increment} $N$ 
		 \STATE $t = t + 1$
		 \ENDFOR
		 \STATE \textbf{Output:} $s_{1}s_{2}\cdots s_{n} $ 
		
 	\end{algorithmic}
 \caption{Embedding procedure for multiple bits}
 \label{Embedding procedure of the protocol for several bits}
 \end{algorithm}
 
 \begin{algorithm}[h]
 \textbf{Procedure} $\mathcal{S}.Extract$ 
 	\begin{algorithmic}[1]
		\STATE \textbf{Input}: \textit{Key} $K$, \textit{Synchronized} $N$, \textit{Mails address1} $s_{1}s_{2} \cdots s_{n_{1}}$, \textit{Mails address2} $s_{1}^{'}s_{2}^{'} \cdots s_{n_{2}}^{'}$ 
		 \STATE $s =$ merge the two mails sequences sorted in ascending order of sent date. 
		 \FOR{$ i = 1 \cdots n$}	
		 	 \STATE $m_{i} = \mathcal{S}.ExtractOneBit(K,N,s_{i})$
		 \STATE \textbf{Increment} $N$ 
		 \ENDFOR
		 \STATE \textbf{Output}: $m_{1}||m_{2}||\cdots||m_{n} $ 
 	\end{algorithmic}
 
 \caption{Extracting procedure for multiple bits}
 \label{Extracting procedure of the protocol for several bits}
 \end{algorithm}
 
 The sequences $s_{1}s_{2} \cdots s_{n_{1}}$ and $s_{1}^{'}s_{2}^{'} \cdots s_{n_{2}}^{'}$ are the mails contained respectively in the mail boxes $address1$ and $address2$.
 
\paragraph*{Run-time:}
 The concealment algorithm runs in $\mathit{O}(n)$. Concerning the extraction algorithm, note that mails received in a mail box are recovered in the sending order. This means that the two sequences are sorted by date of mailing. The statement 2 of the extraction algorithm is there to merge these two sequences in a sorted sequence.
 This operation is performed in at most $\mathit{O}(n)$. Thus, the extraction algorithm runs in $\mathit{O}(n)$ + $\mathit{O}(n)$ which is equal to $\mathit{O}(n)$.

 \subsection{Steganography Protocol 2}
 \subsubsection{Secret steganography to state for a bit}
 The idea of this protocol is to do a broadcast with destination addresses. The order of the addresses in the broadcast is very important. This order determines whether the recipient must take the extracted secret message bit of the document or its complement. From the two recipient addresses, one can have only two types of diffusion address with respect of to the order. So $\mathbb{A} = \lbrace <address1, address2>, <address2, address1> \rbrace $. Let $\mathcal{A} $ be a uniform distribution on $\mathbb{A} $.
 We only need to consider mails sent one address recipient's to extract the secret message.
 
 \begin{algorithm}[h]
 \textbf{Procedure} $\mathcal{S}'.EmbedOneBit$ 
 	\begin{algorithmic}[1]
		\STATE \textbf{Input}: \textit{Key} $K$, \textit{Hiddentext} $m \in \lbrace 0,1 \rbrace$, \textit{Synchronized} $N$, \textit{history} $h$, \textit{Sent date} $t$ 
		 \STATE $ d \leftarrow \mathbf{S}(h)$ \\
		 \IF{$(F_{K}(N,d) = m)$ } 
		 \STATE $s = (d,<address1,address2>,t)$ 
		 \ELSE
		 \STATE $s = (d,<address2, address1>,t)$
		 \ENDIF
		 \STATE\textbf{Increment} $N$ \\
		 \STATE $h = h||d$ \\
		 \STATE \textbf{Output:} $s$
 	\end{algorithmic}
 
 \caption{Embedding procedure for one bit}
 \label{Embedding procedure of the protocol for one bit}
 \end{algorithm}
 
 \begin{algorithm}[h]
 \textbf{Procedure} $\mathcal{S}'.ExtractOneBit$ 
 	\begin{algorithmic}[1]
 		 \STATE \textbf{Input:} \textit{Key} $K$, \textit{Synchronized} $N$, \textit{mail} $s$ 
 		 \STATE $d =$ extractDocument($s$)
 		 \STATE $ m = F_{K}(N,d)$ 
 		 \IF{$(extractAddresses(s) = \hspace*{0.2cm} <address2, address1>) $} 
 		 \STATE $m = \overline{m}$
 		 \ENDIF
 		 \STATE \textbf{Increment} $N$ \\
 		 \STATE \textbf{Output:} $m$ 
 	\end{algorithmic}

 \caption{Extracting procedure for one bit}
 \label{Extracting procedure of the protocol for one bit}
 \end{algorithm}
 
\paragraph*{Run-time:}
 The execution time of this protocol is practically constant for concealment and extraction algorithms and depends on the size of the secret message that is obviously 1: $T = \mathit{O}(1)$.
 
\paragraph*{Insecurity:}
 The security of this protocol is exactly the same as the one expressed in the theorem 1.

 \subsubsection{Secret steganography state for several bits}
 
 \begin{algorithm}[h]
 \textbf{Procedure} $\mathcal{S}'.EmbedMultiBits$ 
 	\begin{algorithmic}[1]
		\STATE \textbf{Input}: \textit{Key} $K$, \textit{Hiddentext} $m \in \lbrace 0,1 \rbrace ^{*}$,\textit{Synchronized} $N$, \textit{history} $h$, \textit{Sent date} $t$ 
		 \STATE \textbf{Parse} $m$ \textbf{as} $m_{1}^{1}||m_{2}^{1}||\cdots||m_{n}^{1} $
		 \FOR{$ i = 1 \cdots n$} 
		 \STATE $s_{i} = \mathcal{S}'.EmbedOneBit(K,m_{i},N,h,t)$ 
		 \STATE $d_{i} = extractDocument(s_{i})$
		 \STATE $h = h||d_{i}$ 
		 \STATE \textbf{Increment} $N$ 
		 \STATE $t = t + 1$
		 \ENDFOR
		 \STATE \textbf{Output:} $s_{1}s_{2}\cdots s_{n} $
 	\end{algorithmic}
 
 \caption{Embedding procedure for multiple bits}
 \label{Embedding procedure of the protocol for several bits}
 \end{algorithm}
 
 \begin{algorithm}[h]
 \textbf{Procedure} $\mathcal{S}'.ExtractMultisBits$ 
 	\begin{algorithmic}[1]
		\STATE \textbf{Input}: \textit{Key} $K$, \textit{Synchronized} $N$, \textit{Mails } $s_{1}s_{2} \cdots s_{n}$ 
		 \FOR{$ i = 1 \cdots n$} 
		 \STATE $m_{i} = \mathcal{S}'.ExtractOneBit(K,N,s_{i})$ 
		 \STATE \textbf{Increment} $N$ 
		 \ENDFOR
		 \STATE \textbf{Output}: $m_{1}||m_{2}||\cdots||m_{n} $ 
 	\end{algorithmic}
 
 \caption{Extracting procedure for multiple bits}
 \label{Extracting procedure of the protocol for several bits}
 \end{algorithm}
 
% \newpage
\paragraph*{Run-time:} $T = \mathit{O}(n)$.
 
 \section{Conclusion}
 In this work, we were interested in the reducing of the time complexity and the number of documents sent to hide a secret message. Our protocols use a constant time to hide and extract a bit of secret message and run in $\mathit{O}(n)$ to hide and extract $n$ bits of a secret message.
 
 \section*{Acknowledgements}
 This work was supported by {\it UMMISCO }, by {\it LIRIMA } and by the {\it University of Yaounde 1}.

\end{document}